\documentclass[aps,pra,showpacs,twocolumn]{revtex4-1} 

\usepackage{graphicx}
\usepackage{dcolumn}
\usepackage{bm}
\usepackage{color}
\usepackage{amsmath}
\usepackage{amssymb}

\begin{document}
\title{Spectrally pure states at telecommunications wavelengths from periodically poled $M$TiO$X$O$_4$ ($M$ = K, Rb, Cs; $X$ = P, As) crystals}
\author{Rui-Bo Jin$^{1}$}
\author{Pei Zhao$^{1}$}
\author{Peigang Deng$^{1}$}
\author{Qing-Lin Wu$^{2}$}
\affiliation{$^{1}$School of Science and Laboratory of Optical Information Technology, School of Material Science and Engineering, Wuhan Institute of Technology, Wuhan 430205, China}
\affiliation{$^{2}$Department of Physics, Central China Normal University, Wuhan 430079, China}

\date{\today }

\begin{abstract}
Significant successes have recently been reported in the study of the generation of spectrally pure state in group-velocity-matched (GVM) nonlinear crystals. However, the GVM condition can only be realized in limited kinds of crystals and at limited wavelengths.
Here, we investigate pure state generation in the isomorphs of PPKTP crystal: i.e., periodically poled RTP, KTA, RTA and CTA crystals. By numerical simulation, we find that these crystals from the KTP family can generate pure photons with high spectral purity (over 0.8), wide tunability (more than 400 nm), reasonable  nonlinearity at a variety of wavelengths (from 1300 nm to 2100 nm). It is also discovered that the PPCTA crystal may achieve purity of 0.97 at 1506 nm. This study may provide more and better choices for quantum state engineering at telecom wavelengths.
\end{abstract}



\maketitle

\section{\emph{Introduction}}
The generation of single photons is a fundamental resource required for optical quantum information processing (QIP) and
spontaneous parametric down conversion (SPDC) is one of the most widely used methods to prepare single photons.
In the general case of the SPDC process, a pump photon is split into two lower energy daughter photons, the signal and idler, which are spectrally correlated.
For many QIP applications, however, it is necessary to utilize bi-photons with no spectral correlations, so as to achieve high-visibility interference between independent sources \cite{Mosley2008, Jin2011,  Jin2013PRA}.

There are two methods to remove the spectral correlations between the bi-photons in SPDC. The first one is to filter the bi-photons using narrow bandpass filters, which can be performed easily, but inevitably and severely decreases the brightness of the photon source. The second method is to engineer the SPDC process, so as to prepare an intrinsically spectrally pure state. Such a quantum state engineering method can be realized by considering the group-velocity-matched (GVM) condition in specific  crystals at certain wavelengths \cite{Grice2001,  Konig2004, Edamatsu2011}. Previous work in the field has shown that the GVM condition can be realized in several crystals, e.g., KDP crystal at 830 nm \cite{Mosley2008, Jin2011, Jin2013PRA2}, periodically poled KTP crystal (PPKTP) at 1584 nm \cite{Konig2004, Jin2015OE} and BBO crystal at 1514 nm \cite{Grice2001, Lutz2013OL, Lutz2014}.

Recently, the study of  spectrally pure state generation in a  PPKTP crystal has achieved significant progress \cite{ Evans2010, Gerrits2011, Eckstein2011, Jin2013OE, Zhou2013, Bruno2014OE}. Besides high spectral purity, the PPKTP crystal still have several other merits, such as  high brightness (because the crystal can be very long, thanks to the periodically poling technique), high damage threshold (higher than PPLN crystal), wide tunability (more than 200 nm at telecom wavelengths \cite{Jin2013OE}).

However, as mentioned above, the GVM condition can only be realized in these limited crystals and at limited wavelengths.
Is it possible to find more crystals to satisfy the GVM condition and can this be done over a wider range of  wavelengths with higher nonlinear efficiency and higher purity? In this work we answer this question by investigating the spectral purity of the photons generated from the isomorphs of a PPKTP crystal: i.e., periodically poled RTP ($\mathrm{RbTiOPO_4}$), KTA ($\mathrm{KTiOAsO_4}$), RTA ($\mathrm{RbTiOAsO_4}$) and CTA ($\mathrm{CsTiOAsO_4}$). We expect these crystals will retain the same highly spectral purity with wide tunability as PPKTP, and possibly provide higher nonlinearity and purity than PPKTP crystal. This study may suggest a wider range of choices for the crystals employed when conducting quantum state engineering at telecom wavelengths.

\begin{table*}[tbp]
\centering\begin{tabular}{c|ccccc}
\hline \hline
Name &PPKTP               &PPRTP                 &PPKTA                 &PPRTA                  &  PPCTA    \\
Composition &$\mathrm{KTiOPO_4}$ &$\mathrm{RbTiOPO_4}$  &$\mathrm{KTiOAsO_4}$  &$\mathrm{RbTiOAsO_4}$  &$\mathrm{CsTiOAsO_4}$ \\
 \hline
$\lambda_\textrm{GVM}$ (nm)    &1584   &1643.2                &1634.7                 &1784.5                 &1864.6 \\
$\Lambda$ ($\mu$m)       &46.1   &56.6                  &57.3                   &71.1                   &381.9\\
Purity                &0.82   &0.82                  &0.82                   &0.82                   &0.82\\
$d_\textrm{eff}$ (pm/V)         &2.4    &2.4                   &2.3                    &2.4                    &2.1 \\
References              &\cite{Konig2004, Fradkin1999,  Kato2002}    &\cite{Mikami2009}                   &\cite{Kato1994}                    &\cite{Cheng1994}                      &\cite{Cheng1993}  \\
\hline \hline
\end{tabular}
\caption{\label{table1} Comparison of the chemical composition, group-velocity-matched (GVM) wavelength $\lambda_\textrm{GVM}$, poling period $\Lambda$, purity and  effective nonlinear coefficient $d_\textrm{eff}$  of PPKTP crystal and four of its isomorphs. The $d_\textrm{eff}=d_{32}$ values are taken from the  SNLO v66 software package, developed by AS-Photonics, LLC \cite{SNLO66}.  The relevant sources in the literatures for the appropriate Sellmeier equations are also listed in the table. }
\end{table*}

\section{Basic Parameters of the four crystals}
The four crystals (RTP, KTA, RTA and CTA) considered in the work are the isomorphs of the KTP crystal, i.e., they belong to the ``KTP'' family. They have the general form of $M$TiO$X$O$_4$ with \{$M$ = K, Rb, Cs\} and \{$X$ = P, As (for $M$ = Cs only)\} \cite{Cheng1994}. Therefore, they retain all the general  properties of their better-known ``parent'' KTP crystal \cite{Dmitriev1999}. For example, they are all positive biaxial crystals (point group: mm2);    they have a long transparency range (0.35 $\mu$m - 3 $\mu$m), reasonable nonlinearity, reduced walk-off in the X-Y plane and  a high optical damage threshold. All these crystals possess ferroelectric properties, and are suitable for use in  periodically polled structures \cite{Shur2016, Karlsson1999, Arvidsson1997}.

The GVM condition is the prerequisite for engineering a spectrally pure state, therefore, the GVM wavelength is a key parameter for these crystals.
In the case of Type II SPDC with a PPKTP crystal, the following GVM condition is met at a wavelength of  1584 nm \cite{Konig2004, Jin2015OE, Fradkin1999,  Kato2002}.
\begin{equation}\label{eq1}
2V^{-1}_{g,p}(\lambda/2)=V^{-1}_{g,s}(\lambda)+V^{-1}_{g,i}(\lambda),
\end{equation}
where $V^{-1}_{g,\mu} (\mu=p,s,i)$ is the inverse of the group velocity  $V_{g,\mu}$ for the pump $p$, the signal $s$,  and the idler $i$. $\lambda$ is the degenerate wavelength of the signal and idler.

For RTP, KTA, RTA and CTA crystals, the GVM wavelengths are 1643 nm, 1635 nm, 1785 nm and 1865 nm, respectively, as listed in Tab. \ref{table1}.
These GVM wavelengths are calculated based on their Sellmeier equations.
All these four crystals are suitable for periodically poling, and their poling periods at the GVM wavelength are shown in Tab. \ref{table1}.
The effective nonlinearity $d_\textrm{eff}$ of the isomorphs (from 2.4 pm/V to 2.1 pm/V) is comparable to that of the PPKTP crystal (2.4 pm/V).
%
%

\section{Pure state generation in the isomorphs}
By satisfying the GVM condition, it is possible to prepare spectrally pure bi-photon state using the isomorphs of PPKTP crystal. The spectral purity can be quantitatively evaluated by considering the spectral distribution of the signal and idler photons. This distribution can be described by their joint spectral amplitude (JSA) $f(\omega_1, \omega_2)$, which is the product of the pump envelope function $\alpha (\omega _s +\omega _i )$ and the phase matching function $\phi (\omega _s ,\omega _i)$, i.e., $f(\omega _s ,\omega _i )= \alpha (\omega _s +\omega _i ) \phi (\omega _s ,\omega _i)$.
By applying Schmidt decomposition to the JSA, we can obtain
\begin{equation}\label{eq2}
f(\omega_1, \omega_2)= \Sigma_j c_j
\phi_j(\omega_1)\varphi_j(\omega_2),
\end{equation}
where  ${ \phi_j(\omega_1)  }$ and ${ \varphi_j(\omega_2)  }$ are two orthogonal basis sets of spectral functions, and ${ c_j }$ is the normalized coefficient.
The spectral purity $p$, a parameter describing the degree of spectral uncorrelation between the signal and idler photons, can be calculated as
\begin{equation}\label{eq3}
p ={ \sum _j c_j ^4}.
\end{equation}

\begin{figure*}[tbp]
\centering\includegraphics[width=16cm]{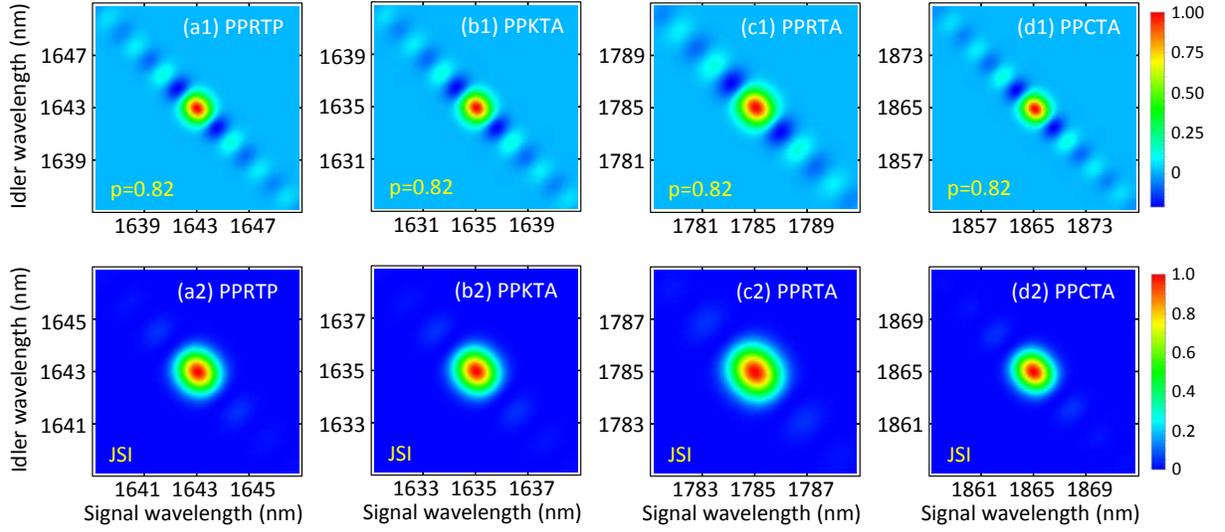}
\caption{(color online)  JSA (first row) and JSI (second row) of the photons generated from the four isomorph crystals at their GVM wavelengths. (a) PPRTP, (b) PPKTA, (c) PRTA and (d) PPCTA. In this simulation, the crystal lengths are fixed at 30 mm long, while the pump laser bandwidths are 0.42 nm, 0.42 nm, 0.50 nm and 0.77 nm for (a), (b), (c) and (d), respectively.
 } \label{isomorph}
\end{figure*}

Under the GVM condition, the photons from a  PPKTP crystal have a high spectral purity of 0.82 at telecom wavelengths.
See Ref.  \cite{Jin2013OE} for more theoretical details of the spectral purity, and the simulation of the  JSA for the PPKTP crystal.
The same property is maintained by the four isomorph crystals studied here.
Figure \ref{isomorph}(a1-d1)  show the JSA $f(\omega _s ,\omega _i )$ of the four crystals at their GVM wavelengths.
In this simulation, we fix the crystal length at 30 mm long  for the four crystals, and scan the pump bandwidth (with a Gaussian spectral distribution) to maximize the spectral purity.
It is noteworthy that all of the four JSA  shown in Fig. \ref{isomorph}(a1-d1) exhibit a  high spectral purity of 0.82 at their GVM wavelengths.
Figure \ref{isomorph}(a2-d2) present the corresponding joint spectral intensities (JSI)  $|f(\omega _s ,\omega _i )|^2$,  which have subcircular shapes.
In this simulation, we assumed a type II ($o \to o + e$) wavelength-degenerated SPDC process in a collinear configuration along the crystal's x-axis.
In the process $o \to o + e$, $o$ indicates polarization along the ordinary (the y-) axis while $e$ denotes polarization along the extraordinary (the z-) axis. The pump and signal photons have polarization along the y-axis, and the idler photons are polarized in the z-axis \cite{Laudenbach2016}.

\section{Wide tunability with high purity in the four crystals}
\begin{figure*}[tbp]
\centering\includegraphics[width=16cm]{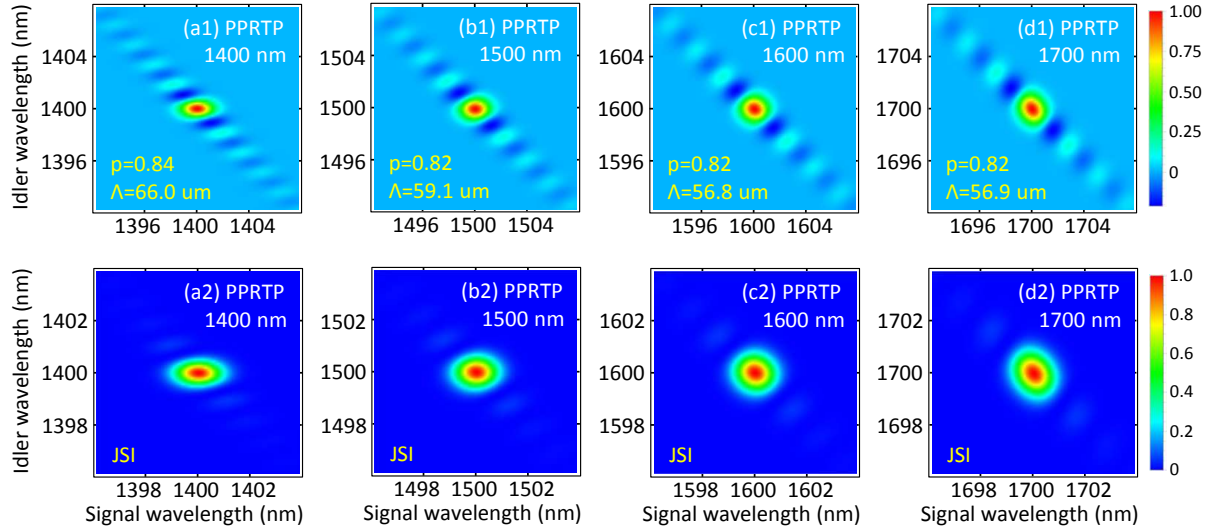}
\caption{(color online)   JSA (first row) and JSI (second row) of photons generated from the PPRTP crystal at different wavelengths. (a)1400 nm, (b)1500 nm, (c)1600 nm, (d)1700 nm. The corresponding spectral purity $p$ and poling period $\Lambda$ are shown in the figure. In this simulation, the crystal lengths are fixed at 30 mm long, while the pump laser bandwidths are fixed at 0.42 nm.
 } \label{PPRTP}
\end{figure*}

At the GVM wavelength, the biphoton source has a high spectral purity and this high purity can be maintained at the nearby wavelengths.
By simulation, we find all these four crystals can maintain wide wavelength tunabilities under high purity.
The purity can remain higher than 0.80,  with wavelength tunable from 1300 nm to 1800 nm for PPRTP crystal; from 1300 nm to 1700 nm for PPKTA crystal; from 1400 nm to 2000 nm for PPRTA crystal; from 1500 nm to 2100 nm for PPCTA crystal.
This range of wavelengths covers the commonly used S-, C-, L-, and U-bands in optical fiber telecommunications.
In Fig. \ref{PPRTP}(a-d) we provide an example of the spectral distribution of the PPRTA crystal at different wavelengths from 1400~nm to 1700~nm  and note that similar property is also possessed by the other three isomorphs and their ``parent'' PPKTP crystal \cite{Jin2013OE}.

\section{Another GVM condition in the CTA crystal}
\begin{figure*}[tbp]
\centering\includegraphics[width=16cm]{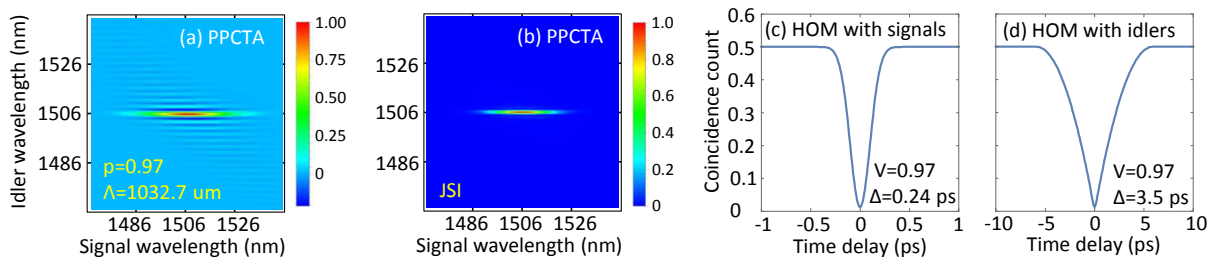}
\caption{ (color online)   (a):  JSA of the photons generated from the PPCTA crystal at 1506 nm, with a purity of 0.97 and poling period of 1032.7 $\mu$m.  (b):  JSI of (a). In this simulation, the crystal lengths are fixed at 30 mm long, while the pump laser bandwidth is 5.0 nm with a center wavelength of 753 nm.  (c):  HOM interference pattern with two heralded signals from (a), with a visibility of 0.97 and width of 0.24 ps. (d): HOM interference pattern with two heralded idlers from (a), with a visibility of 0.97 and width of 3.5 ps. } \label{PPCTA}
\end{figure*}

 While the GVM condition of Eq. (\ref{eq1}) is satisfied in all the isomorphs of PPKTP crystal, a further GVM condition, given by Eq. (\ref{eq4}), is satisfied by the PPCTA crystal at 1506 nm.
\begin{equation}\label{eq4}
V^{-1}_{g,p}(\lambda/2)=V^{-1}_{g,i}(\lambda),
\end{equation}
Under this condition, the JSA can have a very narrow and sharp distribution, as shown in  Fig. \ref{PPCTA}(a).
In this simulation, we assumed a crystal length of of 30 mm for PPCTA and a pump bandwidth of 5 nm  at 753 nm (with a Gaussian profile).
With the JSA in Fig. \ref{PPCTA}(a), the corresponding intrinsical purity is calculated as 0.97, higher than the purities of 0.82 shown in Fig. \ref{isomorph} and Fig. \ref{PPRTP}.
The corresponding JSI is shown in Fig. \ref{PPCTA}(b).
Previously, it was discovered that this condition was satisfied by a KDP crystal at 830 nm \cite{Grice2001, Mosley2008, Jin2011, Jin2013PRA2}.
Here we find, for the first time, that this condition can be satisfied at telecom wavelengths in a PPCTA crystal.

This source offers great potential in quantum interference between independent sources \cite{Mosley2008, Jin2011,  Jin2013PRA, Wu2013}.
In Fig. \ref{PPCTA}(c), we simulate the Hong-Ou-Mandel interference \cite{Hong1987} between two signal photons, one of each from two independent PPCTA crystals (with the JSA shown in Fig. \ref{PPCTA}(a)), with the two idler photons employed as the heralders \cite{Mosley2008, Jin2013PRA, Jin2015OE}. Similarly, Fig. \ref{PPCTA}(d) shows the case of Hong-Ou-Mandel interference between two  idlers heralded by the signal photons. The visibilities  in  Fig. \ref{PPCTA}(c) and (d) are  determined by the the spectral purity shown in  Fig. \ref{PPCTA}(a), which can reach as high as 0.97. The bandwidth of the interference patterns are 0.24 ps and 3.5 ps respectively.

While we only show results for PPCTA here, the conditions of $V^{-1}_{g,p}(\lambda/2)=V^{-1}_{g,i}(\lambda)$ and $V^{-1}_{g,p}(\lambda/2)=V^{-1}_{g,s}(\lambda)$ are also satisfied by the other four crystals. As a results, the similar figure as Fig. \ref{PPCTA}(a) can also be achieved by PPRTP crystal at 1282 nm and 2491 nm, by PPKTA crystal at 1278 nm and 2481 nm, by PPRTA crystal at 1372 nm and 2933 nm, and by PPKTP crystal at 1225 nm and 2337 nm, respectively.

It should be noticed that the JSA in Fig. \ref{PPCTA}(a) is asymmetric, therefore, the Hong-Ou-Mandel interference between the signal and the idler photons (from the same SPDC source) does not show a high visibility, which may limit their use in some quantum information processing protocols.

\section{Future issues}

These results presented here for this theoretical study of these particular isomorph crystals provide a lot of
possible starting points for future theoretical and experimental studies of similar isomorphs and their practical applications to quantum information.
Firstly, in addition to the generation of spectrally decorrelated states shown in   Fig. \ref{isomorph}, it is also possible to prepare spectrally positively-correlated or negatively-correlated states by varying the crystal length or pump bandwidth, in a similar manner to which has been demonstrated in the case of  a PPKTP crystal \cite{Eckstein2011}. Secondly, it is useful to demonstrate customized poiling (by  modulating the  poling-order \cite{Branczyk2011},  the  duty-cycle  \cite{Dixon2013}, or domain-sequence \cite{Dosseva2016, Tambasco2016} ) in these four isomorphs, so as to improve the maximal intrinsically purity from 0.82 to near 1.
Thirdly, it is also meaningful to make waveguide based on these four isomorphs, similar as in the case of PPKTP waveguide \cite{Zhong2009}. Fourthly,  PPCTA crystal shows high spectral purity at around 2 $\mu$m wavelength. This wavelength is useful for biology and medical applications, and also useful for the detection of carbon dioxide \cite{Walsh2009}.  This implies that the quantum state generated from a PPCTA crystal may be a good quantum light source for the quantum information processing in these applications.

In our theoretical model, we only consider the spectral correlation between the signal and idler photons. It is also interesting to consider the spatial correlation, since the spectral purity can be further improved by spatial filtering \cite{Bruno2014OE, Guerreiro2013, Zielnicki2015}. As reported in Ref. \cite{Bruno2014OE}, the purity and coupling efficiency can be improved to around 0.9 by using a larger beam waist for the pump laser, and by coupling the photons into single-mode fibers. But the trade off  is that the source brightness is lower than in the case of tight focusing.

It was reported recently that the KTP family has 118 known isomorphs \cite{Gazhulina2013, Sorokina2007, Stucky1989}, including not only 29 pure crystals but also 89 doped ones. Their general formula can be written as $MM^{\prime}$O$X$O$_4$, where $M$ = K, Rb, Na, Cs, Tl, NH$_4$; $M^{\prime}$ = Ti, Sn, Sb, Zr, Ge, Al, Cr, Fe, V, Nb, Ta, Ga; $X$ = P, As, Si, Ge. The Sellmeier equations for most of these crystals are not reported yet, therefore, exploring the nonlinear optical properties of these crystal is promising for further research. We propose that it is valuable to investigate the GVM wavelength of the pure crystals. For the doped crystals, it may possible to engineer the GVM wavelength to an arbitrary value by adjusting the chemical portions, so as to prepare highly pure photon source at arbitrary wavelength.

\section{Conclusion}
In conclusion, we have theoretically and numerically demonstrated the generation of spectrally uncorrelated states from a small sub-set of the 118 known isomorphs of the PPKTP crystal. It was found that these particular isomorphs  still retain the desirable properties of their parent PPKTP crystal, namely  high spectral purity (over 0.8) with wide tunability (more than 400 nm) at a variety of wavelengths (from 1300 nm to 2100 nm). Further, we found that the PPCTA crystal can achieve an intrinsically high spectral purity of 0.97 at wavelength of 1506 nm. In the future, these crystals may have many promising applications for quantum information processing at telecom wavelengths.

\section*{Acknowledgments}
We thank the anonymous referees for their constructive comments and suggestions to improve our paper. We also acknowledge Robert J. Collins of Heriot-Watt University, Edinburgh, UK for helpful discussions. Rui-Bo Jin is supported by fund from the Educational Department of Hubei Province, China (Grant No. D20161504). Pei Zhao is supported by the National Natural Science Foundation of China (Grant No. 51302199), the Key Natural Science Foundation of Hubei Province of China for distinguished Yong Scholars (Grant No. 2014CFA044) and the Cultivation plan for science and technology talents of WuHan City (Grant No. 2014072704011253). Peigang Deng is supported by the CRSRI Open Research Program (Grant No. CKWV2014225/KY) and the Wuhan Science and Technology Bureau Grant (Grant No. 2015010101010002). Qing-Lin Wu is supported by the Fundamental Research Funds for the Central Universities (CCNU15A02036).


\begin{thebibliography}{99}
\newcommand{\enquote}[1]{#1}

\bibitem{Mosley2008}
P.~J. Mosley, J.~S. Lundeen, B.~J. Smith, P.~Wasylczyk, A.~B. U'Ren,
  C.~Silberhorn, and I.~A. Walmsley, \enquote{Heralded generation of ultrafast
  single photons in pure quantum states,} Phys. Rev. Lett. \textbf{100}, 133601
  (2008).

\bibitem{Jin2011}
R.-B. Jin, J.~Zhang, R.~Shimizu, N.~Matsuda, Y.~Mitsumori, H.~Kosaka, and
  K.~Edamatsu, \enquote{High-visibility nonclassical interference between
  intrinsically pure heralded single photons and photons from a weak coherent
  field,} Phys. Rev. A \textbf{83}, 031805 (2011).

\bibitem{Jin2013PRA}
R.-B. Jin, K.~Wakui, R.~Shimizu, H.~Benichi, S.~Miki, T.~Yamashita, H.~Terai,
  Z.~Wang, M.~Fujiwara, and M.~Sasaki, \enquote{Nonclassical interference
  between independent intrinsically pure single photons at telecommunication
  wavelength,} Phys. Rev. A \textbf{87}, 063801 (2013).

\bibitem{Grice2001}
W.~P. Grice, A.~B. U'Ren, and I.~A. Walmsley, \enquote{Eliminating frequency
  and space-time correlations in multiphoton states,} Phys. Rev. A \textbf{64},
  063815 (2001).

\bibitem{Konig2004}
F.~K{\"o}nig and F.~N.~C. Wong, \enquote{Extended phase matching of
  second-harmonic generation in periodically poled {KTiOPO}$_4$ with zero
  group-velocity mismatch,} Appl. Phys. Lett. \textbf{84}, 1644 (2004).

\bibitem{Edamatsu2011}
K.~Edamatsu, R.~Shimizu, W.~Ueno, R.-B. Jin, F.~Kaneda, M.~Yabuno, H.~Suzuki,
  S.~Nagano, and A. Syouji and K. Suizu, \enquote{Photon pair sources with
  controlled frequency correlation,} Prog. Inform. \textbf{8}, 19--26 (2011).

\bibitem{Jin2013PRA2}
R.-B. Jin, R.~Shimizu, F.~Kaneda, Y.~Mitsumori, H.~Kosaka, and K.~Edamatsu,
  \enquote{Entangled-state generation with an intrinsically pure single-photon
  source and a weak coherent source,} Phys. Rev. A \textbf{88}, 012324 (2013).

\bibitem{Jin2015OE}
R.-B. Jin, T.~Gerrits, M.~Fujiwara, R.~Wakabayashi, T.~Yamashita, S.~Miki,
  H.~Terai, R.~Shimizu, and M.~Takeoka, M. ~Sasaki, \enquote{Spectrally
  resolved {Hong-Ou-Mandel} interference between independent photon sources,}
  Opt. Express \textbf{23}, 28836--28848 (2015).

\bibitem{Lutz2013OL}
T.~Lutz, P.~Kolenderski, and T.~Jennewein, \enquote{Toward a downconversion
  source of positively spectrally correlated and decorrelated telecom photon
  pairs,} Opt. Lett. \textbf{38}, 697--699 (2013).

\bibitem{Lutz2014}
T.~Lutz, P.~Kolenderski, and T.~Jennewein, \enquote{Demonstration of spectral
  correlation control in a source of polarization-entangled photon pairs at
  telecom wavelength,} Opt. Lett. \textbf{39}, 1481--1484 (2014).

\bibitem{Evans2010}
P.~G. Evans, R.~S. Bennink, W.~P. Grice, T.~S. Humble, and J.~Schaake,
  \enquote{Bright source of spectrally uncorrelated polarization-entangled
  photons with nearly single-mode emission,} Phys. Rev. Lett. \textbf{105},
  253601 (2010).

\bibitem{Gerrits2011}
T.~Gerrits, M.~J. Stevens, B.~Baek, B.~Calkins, A.~Lita, S.~Glancy, E.~Knill,
  S.~W. Nam, R.~P. Mirin, R.~H. Hadfield, R.~S. Bennink, W.~P. Grice,
  S.~Dorenbos, T.~Zijlstra, T.~Klapwijk, and V.~Zwiller, \enquote{Generation of
  degenerate, factorizable, pulsed squeezed light at telecom wavelengths,} Opt.
  Express \textbf{19}, 24434--24447 (2011).

\bibitem{Eckstein2011}
A.~Eckstein, A.~Christ, P.~J. Mosley, and C.~Silberhorn, \enquote{Highly
  efficient single-pass source of pulsed single-mode twin beams of light,}
  Phys. Rev. Lett. \textbf{106}, 013603 (2011).

\bibitem{Jin2013OE}
R.-B. Jin, R.~Shimizu, K.~Wakui, H.~Benichi, and M.~Sasaki, \enquote{Widely
  tunable single photon source with high purity at telecom wavelength,} Opt.
  Express \textbf{21}, 10659--10666 (2013).

\bibitem{Zhou2013}
Z.-Y. Zhou, Y.-K. Jiang, D.-S. Ding, and B.-S. Shi, \enquote{{An
  ultra-broadband continuously-tunable polarization-entangled photon-pair
  source covering the C+L telecom bands based on a single type-II PPKTP
  crystal},} J. Mod. Opt. \textbf{60}, 720--725 (2013).

\bibitem{Bruno2014OE}
N.~Bruno, A.~Martin, T.~Guerreiro, B.~Sanguinetti, and R.~T. Thew,
  \enquote{Pulsed source of spectrally uncorrelated and indistinguishable
  photons at telecom wavelengths,} Opt. Express \textbf{22}, 17246--17253
  (2014).

\bibitem{Fradkin1999}
K.~Fradkin, A.~Arie, A.~Skliar, and G.~Rosenman, \enquote{Tunable midinfrared
  source by difference frequency generation in bulk periodically poled
  {KTiOPO}$_4$,} Appl. Phys. Lett. \textbf{74}, 914--916 (1999).

\bibitem{Kato2002}
K.~Kato and E.~Takaoka, \enquote{Sellmeier and thermo-optic dispersion formulas
  for {KTP},} Appl. Opt. \textbf{41}, 5040--5044 (2002).

\bibitem{Mikami2009}
T.~Mikami, T.~Okamoto, and K.~Kato, \enquote{Sellmeier and thermo-optic
  dispersion formulas for {RbTiOPO}$_4$,} Opt. Mater. \textbf{31}, 1628--1630
  (2009).

\bibitem{Kato1994}
K.~Kato, \enquote{Second-harmonic and sum-frequency generation in
  {KTiOAsO}$_4$,} IEEE J. Quantum Electron. \textbf{30}, 881--883 (1994).

\bibitem{Cheng1994}
L.~K. Cheng, L.~T. Cheng, J.~Galperin, P.~A.~M. Hotsenpiller, and J.~D.
  Bierlein, \enquote{Crystal growth and characterization of {KTiOPO}$_4$
  isomorphs from the self-fluxes,} J. Cryst. Growth \textbf{137}, 107--115
  (1994).

\bibitem{Cheng1993}
L.~T. Cheng, L.~K. Cheng, J.~D. Bierlein, and F.~C. Zumsteg, \enquote{Nonlinear
  optical and electro-optical properties of single crystal {CsTiOAsO}$_4$,}
  Appl. Phys. Lett. \textbf{63}, 2618--2620 (1993).

\bibitem{SNLO66}
A.~Smith, \enquote{{SNLO},} {http://www.as-photonics.com/snlo.}

\bibitem{Dmitriev1999}
V.~G. Dmitriev, G.~G. Gurzadyan, and D.~N. Nikogosyan, \emph{Handbook of
  Nonlinear Optical Crystals} (Springer-Verlag Berlin Heidelberg, 1999), 3rd
  ed.

\bibitem{Shur2016}
V.~Y. Shur, E.~V. Pelegova, A.~R. Akhmatkhanov, and I.~S. Baturin,
  \enquote{Periodically poled crystals of {KTP} family: a review,} Ferroelectr.
  \textbf{496}, 49--69 (2016).

\bibitem{Karlsson1999}
H.~Karlsson, F.~Laurell, and L.~K. Cheng, \enquote{{Periodic poling of
  RbTiOPO$_4$ for quasi-phase matched blue light generation},} Appl. Phys.
  Lett. \textbf{74}, 1519--1521 (1999).

\bibitem{Arvidsson1997}
G.~Arvidsson, H.~Karlsson, and F.~Laurell, \enquote{{Periodic poling of
  rubidium titanyl arsenate (RTA) and other crystals from the KTP family},} in
  \enquote{Lasers and Electro-Optics Society Annual Meeting 1997 Conference
  Proceedings., IEEE,} pp. 112--113.

\bibitem{Laudenbach2016}
F.~Laudenbach, H.~Hubel, M.~Hentschel, P.~Walther, and A.~Poppe,
  \enquote{Modelling parametric down-conversion yielding spectrally pure photon
  pairs,} Opt. Express \textbf{24}, 2712--2727 (2016).

\bibitem{Wu2013}
Q.-L. Wu, N.~Namekata, and S.~Inoue, \enquote{High-fidelity entanglement
  swapping at telecommunication wavelengths,} J. Phys. B: At. Mol. Opt. Phys.
  \textbf{46}, 235503 (2013).

\bibitem{Hong1987}
C.~K. Hong, Z.~Y. Ou, and L.~Mandel, \enquote{Measurement of subpicosecond time
  intervals between two photons by interference,} Phys. Rev. Lett. \textbf{59},
  2044--2046 (1987).

\bibitem{Branczyk2011}
A.~M. Bra\'nczyk, A.~Fedrizzi, T.~M. Stace, T.~C. Ralph, and A.~G. White,
  \enquote{Engineered optical nonlinearity for quantum light sources,} Opt.
  Express \textbf{19}, 55--65 (2011).

\bibitem{Dixon2013}
P.~B. Dixon, J.~H. Shapiro, and W.~F.~N. C., \enquote{Spectral engineering by
  {Gaussian} phase-matching for quantum photonics,} Opt. Express \textbf{21},
  5879--5890 (2013).

\bibitem{Dosseva2016}
A.~Dosseva, L.~Cincio, and A.~M. Bra\'nczyk, \enquote{Shaping the joint
  spectrum of down-converted photons through optimized custom poling,} Phys.
  Rev. A \textbf{93}, 013801 (2016).

\bibitem{Tambasco2016}
J.-L. Tambasco, A.~Boes, L.~G. Helt, M.~J. Steel, and A.~Mitchell,
  \enquote{Domain engineering algorithm for practical and effective photon
  sources,} Opt. Express \textbf{24}, 19616--19626 (2016).

\bibitem{Zhong2009}
T.~Zhong, F.~N.~C. Wong, T.~D. Roberts, and P.~Battle, \enquote{{High
  performance photon-pair source based on a fiber-coupled periodically poled
  KTiOPO$_4$ waveguide},} Opt. Express \textbf{17}, 12019--12030 (2009).

\bibitem{Walsh2009}
B.~M. Walsh, \enquote{{Review of Tm and Ho materials; spectroscopy and
  lasers},} Laser Phys. \textbf{19}, 855 (2009).

\bibitem{Guerreiro2013}
T.~Guerreiro, A.~Martin, B.~Sanguinetti, N.~Bruno, H.~Zbinden, and R.~T. Thew,
  \enquote{High efficiency coupling of photon pairs in practice,} Opt. Express
  \textbf{21}, 27641--27651 (2013).

\bibitem{Zielnicki2015}
K.~Zielnicki, K.~Garay-Palmett, R.~Dirks, A.~B. U'Ren, and P.~G. Kwiat,
  \enquote{Engineering of near-ir photon pairs to be factorable in space-time
  and entangled in polarization,} Opt. Express \textbf{23}, 7894--7907 (2015).

\bibitem{Gazhulina2013}
A.~P. Gazhulina and M.~O. Marychev, \enquote{Pseudosymmetric features and
  nonlinear optical properties of potassium titanyl phosphate crystals,} Cryst.
  Struct. Theory App. \textbf{2}, 106--119 (2013).

\bibitem{Sorokina2007}
N.~I. Sorokina and V.~I. Voronkova, \enquote{{Structure and properties of
  crystals in the potassium titanyl phosphate family: A review},} Crystallogr.
  Rep. \textbf{52}, 80--93 (2007).

\bibitem{Stucky1989}
G.~D. Stucky, M.~L.~F. Phillips, and T.~E. Gier, \enquote{The potassium titanyl
  phosphate structure field: a model for new nonlinear optical materials,}
  Chem. Mater. \textbf{1}, 492--509 (1989).

\end{thebibliography}
\end{document}